\documentclass[12pt]{article}

\usepackage[latin1]{inputenc}
\usepackage[english]{babel}
\usepackage{amsmath, amsthm, amsfonts}
\usepackage{epsfig}

\newtheorem{thm}{Theorem}[section]

\newtheorem{lem}[thm]{Lemma}
\newtheorem{prp}[thm]{Proposition}
\newtheorem{hyp}{Hypothesis}
\theoremstyle{definition}
\newtheorem{dfn}[thm]{Definition}
\theoremstyle{remark}
\newtheorem{rem}[thm]{Remark}

\def\RR{\mathbb{R}}

\newcommand{\paren}[1]{\left(#1\right)}
\newcommand{\abs}[1]{\left\vert#1\right\vert}

\author{José A. Cañizo}

\title{Convergence to equilibrium for the discrete
  coagulation-fragmentation equations with detailed balance}

\date{25th May 2007}

\begin{document}
\maketitle

\abstract{Under the condition of detailed balance and some additional
  restrictions on the size of the coefficients, we identify the
  equilibrium distribution to which solutions of the discrete
  coagulation-fragmentation system of equations converge for large
  times, thus showing that there is a critical mass which marks a
  change in the behavior of the solutions. This was previously known
  only for particular cases as the generalized Becker-Döring
  equations. Our proof is based on an inequality between the entropy
  and the entropy production which also gives some information on the
  rate of convergence to equilibrium for solutions under the critical
  mass.}

\section{Introduction}

The discrete coagulation-fragmentation equations (or \emph{DCF
  equations} for short) are a well-known model for physical processes
where a large number of units can join to form groups of two or more.
These equations and their continuous version have been studied
extensively in recent years in the mathematical and physical
literature; as the amount of works dedicated to them is large, we
refer to the classical review \cite{D72} and the more recent
\cite{A99,LM04} for an overall picture of the field, while more
detailed references related to the object of this paper are given
below.

The discrete coagulation-fragmentation equations are:
\begin{equation}\label{eq:DCF}
  \frac{d}{dt} c_j =
  \frac{1}{2} \sum_{k=1}^{j-1} W_{j-k,k} -
  \sum_{k=1}^{\infty} W_{j,k}, \qquad   j \geq 1,
\end{equation}
where
\begin{equation}
  \label{eq:def-Wij}
  W_{i,j} := a_{i,j} c_i c_j - b_{i,j} c_{i+j}
  \qquad i,j \geq 1.
\end{equation}
Here the unknowns are the functions $c_i = c_i(t)$ for $i \geq 1$,
which depend the time $t \geq 0$, and $a_{i,j}, b_{i,j}$ are
nonnegative numbers, the \emph{coagulation} and \emph{fragmentation
  coefficients}, respectively. In the following they are always
assumed to be nonnegative and symmetric in $i,j$. The sum $\sum_{i
  \geq 1} i c_i(t)$ is usually called the \emph{mass of the solution
  at time $t$}, as suggested by the usual physical interpretation of
these equations.

One of the long-standing questions has to do with the long time
behavior of this system, which is expected to model certain phase
change transitions or crystallization processes
\cite{PL79,P97,BCN04,GNON03,NCB02}: under some usual conditions it has
been proved that there is a \emph{critical mass} $\rho_s$ which marks
a qualitative difference in the behavior of the solutions:
\begin{itemize}
\item If a solution $\{c_i\}$ has mass above $\rho_s$, then the
  solution \emph{converges weakly} to the only equilibrium
  distribution with mass equal to $\rho_s$ (meaning that each
  individual $c_i(t)$ converges to the corresponding value for large
  $t$). In this case there is loss of mass in infinite time, as the
  mass of the solution is strictly higher than the mass of the limit
  distribution.
\item If the solution has mass equal to or below $\rho_s$, then it
  \emph{converges strongly} to a unique equilibrium distribution,
  determined by its mass, in the sense that in addition to each $c_i$,
  its average cluster size also converges to its equilibrium value.
  Here mass is also conserved in the limit, as the mass of the
  solution is the same as the mass of the limit distribution.
\end{itemize}

Mathematical proofs of this were first given for the Becker-Döring
system of equations \cite{BCP86,BC88} (which is the particular case of
the DCF equations \eqref{eq:DCF} obtained by setting $a_{i,j} =
b_{i,j} = 0$ whenever both $i$ and $j$ are greater than 1) and then
extended to the generalized Becker-Döring equations
\cite{CdC94,dC98,MR2186002} (obtained by setting $a_{i,j} = b_{i,j} =
0$ whenever both $i$ and $j$ are greater than some fixed $N$). For the
continuous coagulation-fragmentation equations, a proof of weak
convergence to an equilibrium under analogous conditions was given in
\cite{MR1988654}, but to our knowledge there are no available results
on the identification of the concrete equilibrium to which a solution
converges in the continuous setting.

In this paper we show that the same kind of behavior takes place for
the full DCF equations \eqref{eq:DCF} under some conditions on the
coefficients $a_{i,j}$, $b_{i,j}$ which allow, for example, the
following case, which is physically representative \cite{BC90,CdC94}:
\begin{gather}
  a_{i,j} := C(i^\lambda + j^\lambda)\\
  b_{i,j} := C(i^\lambda + j^\lambda)
  \exp \big( C'((i+j)^\mu - i^\mu - j^\mu) \big),
\end{gather}
with $0 \leq \lambda \leq 1$, $0 < \mu < 1$ and some constants $C, C'
> 0$. More explicitly, we show the following:
\begin{thm}
  \label{thm:main-intro}
  Let $c$ be a solution of the DCF equations \eqref{eq:DCF} under
  hypotheses \ref{hyp:growth}--\ref{hyp:moment_1+l_bounded} below;
  call $\rho$ its mass and $\rho_s$ the critical mass.
  \begin{itemize}
  \item If $\rho > \rho_s$ then $c$ converges weakly to the
    only equilibrium with mass $\rho_s$.
  \item If $\rho \leq \rho_s$, then $c$ converges strongly to the only
    equilibrium with mass $\rho$:
  \end{itemize}
\end{thm}
See section \ref{sec:strong-conv-equil} for a more detailed
description of the allowed coefficients, a more complete statement of
the theorem, and its proof. The hypotheses under which we show this
result are specified below as hypotheses
\ref{hyp:growth}--\ref{hyp:moment_1+l_bounded}, and section
\ref{sec:prel-known-results-1} includes a description of the
equilibria. We also prove in section \ref{sec:rate-conv-equil} an
explicit rate of convergence to equilibrium, which is most likely not
optimal and is mainly given as a direct consequence of one of the
inequalities used in the proof:
\begin{thm}
  In the above conditions, if the mass $\rho$ of the solution is
  strictly less than $\rho_s$, then for some constant $C$ depending
  only on the coefficients $a_{i,j}, b_{i,j}$, the mass $\rho$, and
  the moment of order $2-\lambda$ of $c$ at time $t=0$, it holds that
  \begin{equation}
    \sum_{i \geq 1} i \abs{ c_i - c_i^{eq} }
    \leq \frac{C}{\sqrt{1 + \log (1+t)}}
    \quad \text { for all } t > 0,
  \end{equation}
  where $\{c_i^{eq}\}_{i \geq 1}$ is the equilibrium distribution with
  mass $\rho$.
\end{thm}

The main interest of our result is that it identifies the limiting
equilibrium for a general class of coefficients for which all of
$a_{i,j}, b_{i,j}$ are nonzero; as explained above, previous results
in \cite{BCP86,BC88,dC98,CdC94,MR2186002} always imposed that
$a_{i,j}, b_{i,j}$ should be zero whenever both $i$ and $j$ are
greater than a fixed $N$. Our statement extends the corresponding ones
in \cite{BC88,CdC94,MR2186002} except for the fact that we impose a
more restrictive condition on the initial data, namely, that it has a
finite moment of a certain order which is less than two in common
examples. In turn, as explained above, we allow for coefficients in
which none of $a_{i,j},b_{i,j}$ are zero. There are also some
restrictions on the coefficients: for example, our result does not
apply when, for some $C > 0$,
\begin{equation}
  a_{i,j} := C(i^\alpha j^\beta + i^\beta j^\alpha)
\end{equation}
with $\alpha, \beta$ both strictly positive (note that the case
studied in this paper corresponds to $\beta = \lambda$, $\alpha = 0$).
As mentioned above, our main result is based on hypotheses
\ref{hyp:growth}--\ref{hyp:moment_1+l_bounded} below; hypothesis
\ref{hyp:coag_small_particles}, which roughly
states that the strength of small-large interactions is comparable to
that of large-large interactions, does not hold for this $a_{i,j}$,
and whether the result is true also in this case is an open question.

The method of proof of this behavior contained in previous works is
based, as a first step, on the study on an \emph{entropy functional}
for this equation: a quantity which is decreasing along solutions and
allows one to conclude that every solution converges weakly to a
suitable equilibrium; and as a second step, on the development of
estimates on the amount of large particles by means of which one can
identify the average cluster size of the equilibrium to which the
solution converges.

Our proof employs the entropy functional for the second step in a new
way, based on an inequality between it and its derivative, the
\emph{entropy production}, and also a simpler related inequality. As
far as we know, this technique has not been previously employed for
identifying the mass of the equilibrium to which solutions converge.
This approach is inspired in a paper by Jabin and Niethammer
\cite{JN03}, where they prove a similar inequality to study the rate
of convergence to equilibrium of the Becker-Döring equations.  The
argument draws strongly from the entropy-entropy production method
which was successfully employed to study the long-time behavior of the
Aizenman-Bak model \cite{citeulike:1081959} (i.e., the continuous
coagulation-fragmentation equations with constant coefficients), its
inhomogeneous version \cite{CDF} and other kinetic equations, notably
the Boltzmann equation, for which the literature is quite abundant: we
refer the reader to the review \cite{MR2065020} for further references
and background information.

The rough idea is as follows: we use a \emph{relative energy}
functional which we denote by $F_z$ and which, as is well known, is
decreasing along solutions of our equation (this functional plays the
role of entropy here, but is more appropriately named relative energy
in agreement with previous uses and its common physical
interpretation).  Its derivative, which is negative, is called the
\emph{free energy dissipation}, denoted by $D_{CF}$. If the relative
energy functional is chosen appropriately, it measures how far a
solution is from equilibrium; hence, if one can show an inequality
relating $F_z$ and $D_{CF}$, one can deduce from that a differential
inequality for the evolution of $F_z$, and with it an estimate on the
approach of the solution to equilibrium. As mentioned before, this has
been carried out in \cite{JN03} to estimate the rate of convergence to
equilibrium of the Becker-Döring equations.  Concretely, the
inequality proved there states that for a solution with mass below the
critical one and when $c_1$ is less than a certain critical value,
\begin{equation}
  F_z \leq C D \abs{\log \frac{1}{F_z}}^2,
\end{equation}
where $D$ is the dissipation rate (which is just $D_{CF}$ is the
particular case of the Becker-Döring equations) and $C$ is a constant
which depends on the mass of the solution, the distance of $c_1$ to
the critical value mentioned above, and also on a uniform estimate on
certain exponential moments of the solution. If one wants to prove
theorem \ref{thm:main-intro}, the latter estimate is out of reach, as
proving such an estimate would automatically yield the result. Hence,
our intention is to find a weaker inequality which does not require
such a strong estimate of moments of the solution but still allows us
to recover useful information on its long-time behavior. It turns out
that one can find such an inequality: in section
\ref{sec:relat-energy-estim} we show that, under conditions similar to
the ones above but \emph{without assuming boundedness of exponential
  moments}, and for some constant $C$,
\begin{equation}
  F
  \leq
  C \sqrt{D_{CF}} \sqrt{M_{2-\lambda}}
\end{equation}
where $F$ is a slight modification of the relative energy functional (see
section \ref{sec:relat-energy-estim}) and $M_{2-\lambda}$ is the
moment of order $2 - \lambda$ of the solution. An easy estimate on
this moment shows that it increases at most linearly, and this is
enough to deduce theorem \ref{thm:main-intro}. Though one could
actually deduce it this way, instead we use a simpler inequality which
is suggested by this idea and directly gives an estimate on the
difference to the equilibrium mass (see section
\ref{sec:mass-diff-estim}).

Aside from being the key point in the proof of our results on the
asymptotic behavior of the DCF equations, the inequality we prove
below is interesting in itself, as it can be used to deduce other
properties of the coagulation-fragmentation system of equations, or to
obtain stronger inequalities in other cases. As an application, we
give an estimate on the rate of convergence to equilibrium of
solutions with mass below the critical one, which is however by no
means expected to be optimal; in fact, one would expect the solution
to converge to the equilibrium at a rate similar to the one obtained
by Jabin and Niethammer in \cite{JN03}, this is, a convergence like
$e^{-C t^{1/3}}$ for some constant $C > 0$. However, further estimates
on the solution (such as, for example, uniform estimates on
exponential moments) which are not readily available here are
essential in \cite{JN03} in order to show such a convergence. Finding
the optimal rate of convergence for the DCF equations is an
interesting open problem.

In the next section we present some preliminary definitions and known
properties of the solutions. In section \ref{sec:mass-diff-estim} we
prove the inequality which is used in section
\ref{sec:strong-conv-equil} to identify the equilibrium to which a
solution converges, and show our main result on the matter. In section
\ref{sec:relat-energy-estim} we prove the inequality relating the
relative energy to the free energy dissipation rate, and use it to
give an explicit rate of convergence to equilibrium. Finally, in an
appendix (section \ref{sec:H_theorem}) we prove, under conditions
suitable for our result, that the entropy functional is decreasing
along solutions of the DCF equations. This is a result known under
only slightly different conditions, and its proof is given in the
appendix for completeness, though it does not contain essentially new
arguments.

\textbf{Acknowledgements.} I wish to thank Stéphane Mischler and
Clément Mouhot for some discussions concerning these results, which
were extremely helpful. The author was supported by the ANR research
group \emph{SPINADA} and the University Paris-Dauphine.

\section{Preliminaries and known results}
\label{sec:prel-known-results-1}

\subsection{Definitions and hypotheses}

Let us first define precisely what we understand by a solution of the
DCF equations \eqref{eq:DCF}:

\begin{dfn}
  \label{dfn:solution}
  A \emph{solution} on the interval $[0,T)$ (for a given $T > 0$ or
  $T=\infty$) of (\ref{eq:DCF}) is a sequence of nonnegative functions
  $c_i: [0,T) \to [0,+\infty)$ ($i \geq 1$) such that
  \begin{enumerate}
  \item for all $i \geq 1$, $c_i$ is absolutely continuous in compact
    sets of $[0,T)$ and $\sum_{i \geq 1} i c_i(t)$ is bounded on
    $[0,T)$,
  \item for all $j \geq 1$, the sums $\sum_{i=1}^\infty a_{i,j}
    c_i(t)$ and $\sum_{i=1}^\infty b_{i,j} c_{i+j}(t)$ are finite for
    almost all $t \in [0,T)$,
  \item and equations (\ref{eq:DCF}) hold for almost all $t \in [0,T)$.
  \end{enumerate}
\end{dfn}

If $\{c_i\}_{i \geq 1}$ is a solution to the DCF equations on some
interval and $t$ is in this interval, we will refer to the sum
$\sum_{i \geq 1} i c_i(t)$ as its \emph{mass at time $t$}. Actually,
if $\{c_i\}_{i \geq 1}$ is any sequence of nonnegative numbers, we
will call the sum $\sum_{i \geq 1} i c_i$ its mass. We say that a
solution to the DCF equations \emph{conserves the mass} when its mass
at any time of its interval of definition is the same.

An \emph{equilibrium} is a solution of the DCF equations which does
not depend on time.

For the main results in this paper we will need some or all of the
following hypotheses; note that hypotheses
\ref{hyp:growth}--\ref{hyp:coag_small_particles} concern the
coefficients $a_{i,j}, b_{i,j}$, while hypothesis
\ref{hyp:moment_1+l_bounded} concerns the initial data $c^0$.

\begin{hyp}[Growth of coefficients]\label{hyp:growth}
  For integer $i,j \geq 1$, the coefficients $a_{i,j}$ and $b_{i,j}$
  are nonnegative numbers, they are symmetric in $i,j$ (this is,
  $a_{i,j} = a_{j,i}$ and $b_{i,j} = b_{j,i}$ for all $i,j \geq 1$)
  and for some constants $K>0$, $0\leq \lambda <1$ and $\gamma \in
  \RR$,
  \begin{align}
    & a_{i,j}, b_{i,j} \leq K (i^\lambda + j^\lambda)
    \quad
    \text{ for all } i,j \geq 1, \\
    % & b_{i,j} \leq K (i^\lambda + j^\lambda).
    & \sum_{j=1}^{i-1} b_{j,i-j} \leq K i^\gamma
    \quad
    \text{ for all } i \geq 1
  \end{align}
\end{hyp}

\begin{hyp}[Detailed Balance] \label{hyp:db}
  There exists a positive sequence $\{Q_i\}_{i \geq 1}$ with $Q_1=1$
  such that for all $i,j \geq 1$,
  \begin{equation}
    a_{i,j} Q_i Q_j = b_{ij} Q_{i+j}.
  \end{equation}
\end{hyp}

\begin{hyp}[Critical monomer concentration]\label{hyp:Qjk}
  The sequence $Q_i$ satisfies that:
  \begin{equation}
%%    & \left| \log Q_{j+1}- \log Q_j \right| \leq K_Q \quad
%%      \text{for some } K_Q \geq 0 \text{ and all } j \geq 1, \\
%     \label{eq:Qjk1}
%     & \log Q_j + \log Q_k \leq \log Q_{j+k}
%        \quad \text{for all } j,k \geq 1, \\
    \label{eq:Qjk2}
    \lim_{j\to\infty} Q_j^{1/j} = \frac{1}{z_s}
    \quad \text{ for some } 0 < z_s < \infty.
  \end{equation}
  The critical density $\rho_s$ is defined to be
  \begin{equation}
    \rho_s := \sum_{j=1}^\infty j Q_j z_s^j,
    \quad \quad (0 < \rho_s \leq \infty).
  \end{equation}
\end{hyp}

% This implies that
% %   $\lim_{j\to\infty} \frac{Q_j}{Q_{j+m}} = z_s^m$
% %   for $m \geq 1$ and that
% \begin{equation}
%   \lim_{j\to\infty} Q_j^{1/j} = \frac{1}{z_s}.
% \end{equation}

\begin{hyp}[Regularity of $Q_i$]
  \label{hyp:Qkj_regular}
%   It holds that:
%   \begin{enumerate}
%   \item
%     \label{item:Qi_zsi_decreasing}
    The sequence $\{Q_i z_s^i\}_{i \geq 1}$ is decreasing.
%   \item
%     \label{item:Q_i+1/Q_i_bounded}
%     $\left| \log Q_{i+1}- \log Q_i \right| \leq K_Q$
%     for some $K_Q \geq 0$ and all $i \geq 1$.
%   \end{enumerate}
\end{hyp}

\begin{hyp}[Strong coagulation of small particles]
  \label{hyp:coag_small_particles}
  For some constant $K_1 > 0$ it holds that
  \begin{equation}
    a_{i,1} \geq K_1 i^\lambda
    \quad
    \text{ for all } i \geq 1.
  \end{equation}
\end{hyp}

\begin{hyp}[Moment of initial data]
  \label{hyp:moment_1+l_bounded}
  The sequence $\{c_i^0\}_{i \geq 1}$ (which will be used as initial
  data later) is a sequence of nonnegative numbers with finite moments
  of orders $2-\lambda$, $1+\lambda$ and $1+\gamma$; this is,
  \begin{equation}
    \sum_{i \geq 1} i^\mu c_i^0 < +\infty
    \quad \text{ for }
    \mu := \max \{ 2-\lambda, 1+\lambda, 1+\gamma \}.
  \end{equation}
\end{hyp}

% \renewcommand{\labelenumi}{H\Alph{enumi}}
% \begin{enumerate}
% \item
%   \label{l1}fhdsjkf
%   fhdjskfh
% \item   \label{l2}fjdskfh
% \item hfsdjkhfs
% \end{enumerate}
% Referencia \ref{l2}.

\subsection{Existence of solutions and equilibria}
\label{sec:existence-solutions}

Next we recall some known results on the existence, uniqueness and
properties of solutions of the DCF equations.

In order to derive estimates on solutions it is often useful to see
them as a limit of solutions of simpler systems.  In this case it is
common to consider the finite system of ordinary differential
equations obtained by taking an $N \geq 1$ and writing the DCF
equations with coefficients $a_{i,j}^N$, $b_{i,j}^N$, where
\begin{align}
  a_{i,j}^N = a_{i,j}, \quad b_{i,j}^N = b_{i,j} &\quad \text{ for }
  i+j \leq N
  \\
  a_{i,j}^N = b_{i,j}^N = 0 &\quad \text{ for } i+j > N,
\end{align}
and taking into account equations for $c_i$ only up to $i = N$,
while $c_i$ are taken to be 0 for $i > N$. For any nonnegative
initial data $\{c_i^0\}_{i \leq N}$ at $t=0$ this finite system is
shown to have a unique nonnegative solution defined on $[0,+\infty)$
\cite{BC90}.

Existence results are usually obtained by proving that the sequence of
truncations just defined converges in some sense and its limit is a
solution of the complete DCF equations. Let us state a result of this
kind taken from Theorems 2.4 and 2.5 and Corollary 2.6 of \cite{BC90}:
\begin{prp}[Existence of solutions]
  \label{prp:existence_discrete}
  Assume hypothesis \ref{hyp:growth}, and take any nonnegative
  sequence $\{c_i^0\}_{i \geq 1}$ with $\sum_{i \geq 1} i c_i^0 <
  +\infty$. Then there exists a mass-conserving solution $c$ to the
  DCF equations (\ref{eq:DCF}) on $[0, +\infty)$ with $c(0) = c^0$.

  In addition, this solution is constructed as a limit of solutions of
  the truncated system defined at the beginning of section
  \ref{sec:existence-solutions} in the sense that, if $\{c_i^N\}_{i
    \leq N}$ is the solution of the finite truncated system with $N$
  equations and initial data $\{c_i^0\}_{i \leq N}$, then there is
  some sequence $\{N_k\}_k$ such that for all $T > 0$
  \begin{equation}
    \underset{t \in [0, T)}{\sup}
    \sum_{i \geq 1} i \abs{c_i(t) - c_i^{N_k}(t)}
    \to 0
    \quad \text{ when } k \to \infty.
  \end{equation}
  Note that $c_i^N$ is taken to be 0 whenever $i > N$.
\end{prp}

% A useful uniqueness result is found in \cite[Corollary 5.2]{L02}. Note
% that the multiple fragmentation case studied in this paper has the
% system \eqref{eq:DCF} as a particular case:
% \begin{prp}[Uniqueness of the solution]
%   Assume hypothesis \ref{hyp:growth} on the coefficients and take
%   nonnegative initial data $c^0 = \{c_i^0\}_{i \geq 1}$ such that
%   $\sum_{i \geq 1} i^{1+\lambda} c_i^0$ is finite. Then there is one
%   and only one solution $c = \{c_i\}_{i \geq 1}$ to the DCF equations
%   with $c(0) = c^0$ which conserves the mass and satisfies, for all $T
%   > 0$, that
%   \begin{equation}
%     \underset{t \in [0,T]}
%     \sup \sum_{i \geq 1} i^{1 + \lambda} c_i(t) < \infty.
%   \end{equation}
% \end{prp}

The following result on the existence of equilibria can be found in
\cite[Theorem 5.2]{CdC94}:
\begin{prp}
  \label{prp:equilibria}
  Assume hypotheses \ref{hyp:growth}--\ref{hyp:Qjk}.
  \begin{enumerate}
  \item For $0 \leq \rho \leq \rho_s$ (or $\rho < +\infty$ if $\rho_s
    = +\infty$), there exists exactly one equilibrium $\{c_i^\rho\}_{i
      \geq 1}$ of (\ref{eq:DCF}) with density $\rho$, which is given
    by
    \begin{equation}
      c_i^\rho = Q_i z^i \quad \forall \, i \geq 1,
    \end{equation}
    where $0 \leq z \leq z_s$ is the only positive number such that
    $\sum_{i=1}^\infty i Q_i z^i = \rho$.
  \item For $\rho_s < \rho < +\infty$ there is no equilibrium of
    (\ref{eq:DCF}) with density $\rho$.
  \end{enumerate}
\end{prp}

\subsection{Lyapunov functionals}
\label{sec:Lyapunov}

Take $\{Q_i\}$ to be the sequence defined in hypothesis \ref{hyp:db};
we assume hypothesis \ref{hyp:Qjk} throughout. If $c = \{c_i\}$ is a
nonnegative sequence with finite mass ($\sum_{i \geq 1} i\,c_i <
+\infty$), then we define the \emph{free energy} $V(c)$ as:
\begin{equation}
  \label{eq:free-energy}
  V(c)
  := \sum_{i=1}^\infty
  c_i  \paren{
    \log \frac{c_i}{Q_i}
    - 1
  }
\end{equation}
(Observe that hyp. \ref{hyp:db} ensures that $Q_i > 0$ for all $i$.)
When $c$ is understood we will simply denote this as $V$. In
\cite[Lemma 4.2 and p. 680]{BCP86} it is proved that it is finite for
all nonnegative $c$ with finite mass and that, for any $\rho \geq 0$,
it is bounded above and below on the set of nonnegative sequences
$\{c_i\}$ such that $\sum_{i \geq 1} i\, c_i = \rho$ (always under
hypothesis \ref{hyp:Qjk}; see also lemmas \ref{lem:moment_k+logc} and
\ref{lem:moment_k+logQ}).

% Let $\{c_i\}_{i \geq 1}$ be a (nonzero) mass-conserving solution of
% the coagulation-fragmentation equations on $[0,+\infty)$, and denote
% its total mass by $\rho$. We know that it converges weak-* to some
% equilibrium given by ${Q_i z^i}_{i \geq 1}$ for some $0 < z \leq z_s$.

If $0 < z \leq z_s$ and $c$ is as above, we define the \emph{free
  energy relative to the equilibrium $\{Q_i z^i\}_{i \geq 1}$}, or
\emph{relative energy} for short, by the following expression, after
Jabin and Niethammer \cite{JN03}:
\begin{align}
  F_z(c)
  & := \sum_{i=1}^\infty
  c_i  \paren{
    \log \frac{c_i}{Q_i z^i}
    - 1
  }
  + \sum_{i=1}^\infty Q_i z^i
  \\
  & = V(c)
  + \sum_{i=1}^\infty Q_i z^i
  - \log z \sum_{i \geq 1} i\, c_i
\end{align}
where $c^{eq}$ represents the equilibrium $\{Q_i z^i\}_{i \geq 1}$.
This is also clearly finite when $0 < z < z_s$; for $z = z_s$ it is
finite when $\rho_s < +\infty$, but may be infinite when $\rho_s =
+\infty$. Also, when the mass of $c$ is finite and less than or equal
to $\rho_s$, we can choose $z$ so that the mass of the equilibrium
$c^{eq} := \{Q_i z^i\}_{i \geq 1}$ is the same as that of $c$. In this
case, $F_z(c)$ can be written as the difference between the free
energy of $c$ and that of the equilibrium with the same mass:
\begin{equation}
  \label{eq:F_z-as-difference-of-free-energy}
  F_z(c) = V(c) - V(c^{eq}).
\end{equation}

Finally, when $c_i > 0$ for all $i$, the \emph{free energy dissipation
  rate} $D_{CF}(c)$ is defined as
\begin{equation}
  \label{eq:dissipation_rate}
  D_{CF}(c)
  :=
  \frac{1}{2}
  \sum_{i,j= 1}^\infty
  a_{i,j} Q_i Q_j
  \paren{
    \frac{c_i c_j}{Q_i Q_j} - \frac{c_{i+j}}{Q_{i+j}}
  }
  \paren{
    \log \frac{c_i c_j}{Q_i Q_j} - \log \frac{c_{i+j}}{Q_{i+j}}
  }
  \geq
  0.
\end{equation}

Now, assume hypotheses \ref{hyp:growth}--\ref{hyp:Qjk} and also that
for all $i \geq 1$, $a_{i,1} > 0$ (which implies $b_{i,1} > 0$). Let
$\{c_i\}_{i \geq 1}$ be the solution to the DCF equations
\eqref{eq:DCF} on $[0, +\infty)$ given by proposition
\ref{prp:existence_discrete} under these hypotheses. The positivity
assumption on $a_{i,1}$ above ensures that $c_i(t) > 0$ for all $t >
0$ and $i \geq 1$ (see \cite{C92} or \cite[Theorem 5.2]{CdC94}), so
$D_{CF}(c(t))$ makes sense for $t > 0$. Denote $V \equiv V(c(t))$,
$F_z \equiv F_z(c(t))$ and $D_{CF} \equiv D_{CF}(c(t))$ for $t \geq
0$. We prove in section \ref{sec:H_theorem} that if the initial
condition has certain finite moments, then both $V$ and $F_z$ are
absolutely continuous on compact sets and
\begin{equation}
  \frac{d}{dt} F_z = \frac{d}{dt} V = -D_{CF}
  \quad
  \text{ for almost all } t > 0.
\end{equation}

Then, $V$ and $F_z$ are decreasing along mass-conserving solutions of
\eqref{eq:DCF}, so they are Lyapunov functionals for this equation
(they differ by a constant along a given solution). We will be
especially interested in studying $F_z$, as it is positive and measures
the proximity of a solution to the equilibrium $\{Q_i z^i\}$ in the
following sense, taken from \cite[Lemma 3.1]{JN03}: if $c = \{c_i\}_{i
  \geq 1}$ is a nonnegative sequence with finite mass and $0 < z <
z_s$, then there is a constant $K_z$ such that
\begin{equation}
  \label{eq:F_bounds_strong_convergence}
  \sum_{i \geq 1} i\, \abs{c_i - Q_i z^i}
  \leq
  \max \left\{
    2 F_z(c), K_z \sqrt{F_z(c)}
  \right\}.
\end{equation}
The constant $K_z$ can be taken to be
\begin{equation}
  K_z := \frac{1}{1-\sqrt{\mu}} - 1
  \quad \text{ with } \mu := \frac{z}{z_s}.
\end{equation}
A consequence of this is that, for a given mass-conserving solution
$c$ of \eqref{eq:DCF} with mass $\rho$, $F_z(c(t)) \to 0$ as $t \to
+\infty$ for some $0 < z < z_s$ implies that $z$ is such that $\sum_{i
  \geq 1} i\, Q_i z^i = \rho$.
% Actually, later we will need some
% control on the constant $K_z$, which can be obtained by following the
% proof in Lemma 3.1 of \cite{JN03} (one can actually get an explicit
% expression for $K_z$ and see that $K_z \to 0$ as $z \to 0$). However,
% the following result will be enough for our purposes and is
% interesting in itself:

% \begin{lem}
%   \label{lem:loss_of_mass_bounded_by_F}
%   For any nonnegative sequence $c = \{c_i\}_{i \geq 1}$ and any $0 < z
%   < z_s$ it holds that
%   \begin{equation}
%     \sum_{i \geq 1} i c_i - \sum_{i \geq 1} i Q_i z^i
%     \leq
%     \frac{1}{\log\frac{z_s}{z}}
%     F_z(c).
%   \end{equation}
% \end{lem}

Below we will use the free energy dissipation rate which appears in
the Becker-Döring equations: for a strictly positive sequence
$\{c_i\}_{i \geq 1}$ we set
\begin{equation}
  \label{eq:BD_dissipation}
  D
  :=
  \sum_{i = 1}^\infty
  a_i Q_i
  \paren{
    \frac{c_1 c_i}{Q_i} - \frac{c_{i+1}}{Q_{i+1}}
  }
  \paren{
    \log \frac{c_1 c_i}{Q_i} - \log \frac{c_{i+1}}{Q_{i+1}}
  }
\end{equation}
where
\begin{align}
  \label{eq:def-a_1}
  &a_1 := \frac{1}{2} a_{1,1},\\
  \label{eq:def-a_i}
  &a_i := a_{i,1} \quad \text{ for } i > 1,
\end{align}
Note that $0 \leq D \leq D_{CF}$, as every term in $D$ already appears
in $D_{CF}$, taking into account the symmetry of $a_{i,j}$.

\subsection{H-theorem}

As was already pointed out above, formally one can calculate the time
derivative of the free energy to obtain that
\begin{equation}
  \frac{d}{dt} V = -D_{CF}.
\end{equation}
This result has been proved rigorously in \cite[Theorem 5.2]{CdC94}
under a growth hypothesis on the coefficients $a_{i,j}$, $b_{i,j}$ and
some further regularity assumptions given as conditions on the
sequence $Q_i$; for the continuous equations, the corresponding result
was proved in \cite{MR1988654} by assuming a stronger regularity of
the initial condition (namely, the boundedness of certain moments) and
comparatively weaker regularity of the coefficients $a_{i,j}$,
$b_{i,j}$. Here we would like to prove the result for the discrete
equations in a way similar to that in \cite{CdC94}, but which uses
hypotheses analogous to those in \cite{MR1988654}. This result is more
natural in our context, as anyway an essential point of the proof of
our main result relies on moment estimates for the solution.
\begin{thm}
  \label{thm:H}
  Assume hypotheses \ref{hyp:growth}--\ref{hyp:Qjk} and also that
  \begin{enumerate}
  \item the initial data $c^0 = \{c^0_i\}_{i \geq 1}$ is nonnegative
    and has finite moments of order $1+\lambda$ and $1+\gamma$,
  \item $a_{1,i}, b_{1,i} > 0$ for all $i \geq 1$.
  \end{enumerate}
  (We recall that $\lambda$ and $\gamma$ are defined in hypothesis
  \ref{hyp:growth}.)  Let $c$ be the solution of the DCF equations
  given by theorem \ref{prp:existence_discrete}. Then, $D_{CF}$ is
  locally integrable and
  \begin{equation}
    \frac{d}{dt} V = -D_{CF}
    \quad
    \text{ for almost all } t \geq 0.
  \end{equation}
\end{thm}
The proof of this is given in section \ref{sec:H_theorem}, as it is
only a slight variation of well-known proofs such as those in
\cite{MR1988654, CdC94}.

\subsection{Weak convergence of solutions to an equilibrium}

As stated in theorem \ref{thm:H}, the free energy $V$ (defined in eq.
\eqref{eq:free-energy}) is decreasing along solutions of the DCF
equations. It is known that this implies that every solution must
converge in a weak sense to a certain equilibrium with mass less than
or equal to that of the solution itself, as is shown for example in
\cite[Theorem 6.4]{CdC94}. We state this in the following result,
which is known to hold under slightly different hypotheses; its proof
follows from the H-theorem \ref{thm:H} in the same way as in
\cite[Theorem 6.4]{CdC94}:
\begin{prp}
  \label{prp:weak-star-convergence}
  Assume the same hypotheses as in theorem \ref{thm:H}. Let $c =
  \{c_j\}$ be a solution of (\ref{eq:DCF}) on $[0,\infty)$ which
  conserves mass, and call its mass $\rho$.

  Then there exists $0 \leq z \leq z_s$ such that $\sum_{i \geq 1} i\,Q_i
  z^i \leq \rho$ and
  \begin{equation}
    \lim_{t \to +\infty} c_i(t) = Q_i z^i
    \quad \text{ for all } i \geq 1
  \end{equation}
\end{prp}

The above convergence is usually referred to as \emph{weak-$*$
  convergence}. Precisely, we say that a sequence $c^n = \{c^n_i\}_{i
  \geq 1}$ \emph{converges weak-$*$} to a sequence $c = \{c_i\}_{i \geq 1}$ if
\begin{itemize}
\item $\sum_{i \geq 1} i \abs{c^n_i} \leq K$ for some $K > 0$ and all
  $n \geq 1$, and
\item for each $i \geq 1$, $c^n_i \to c_i$ as $n \to \infty$.
\end{itemize}
We denote this as $c^n \overset{*}{\rightharpoonup} c$. There is also
a useful relationship between weak-$*$ and strong convergence:
\begin{lem}[\cite{BCP86}, Lemma 3.3]
  \label{lem:weak-strong-convergence}
  If $\{c^n\}$ is a sequence such that $c^n
  \overset{*}{\rightharpoonup} c$ and $\sum_{i \geq 1} i \abs{c^n_i}
  \to \sum_{i \geq 1} i \abs{c_i}$, then $\sum_{i \geq 1} i \abs{c^n_i
    - c_i} \to 0$.
\end{lem}

\section{Mass difference estimate}
\label{sec:mass-diff-estim}

In this section we prove the following inequality, which is the
fundamental result needed to prove our main result, theorem
\ref{thm:main}:
\begin{prp}
  \label{prp:mass_difference_bound}
  Assume hypotheses \ref{hyp:growth}--\ref{hyp:coag_small_particles}
  and take a strictly positive sequence $c = \{c_i\}_{i \geq 1}$.
  Suppose that
  \begin{itemize}
  \item $0 < c_1 < z_s$ and
  \item $M_{2-\lambda} := \sum_{i \geq 1} i^{2-\lambda} c_i < +\infty$,
  \end{itemize}
  and call $D \equiv D(c)$ the Becker-Döring free energy dissipation
  rate from equation \eqref{eq:BD_dissipation}. Then,
  \begin{equation}
    \sum_{i \geq 1} i\, c_i - \sum_{i \geq 1} i\,Q_i c_1^i
    \leq
    C
    \sqrt{D}
    \sqrt{M_{2-\lambda}}
  \end{equation}
  for some constant $C$ depending only on the coefficients $a_{i,j}$,
  $b_{i,j}$ and increasingly on the quantity $c_1/z_s$.  
\end{prp}
The lemma which follows will be used in the proof of this result:
% and the second one is a simple analysis result needed to prove the
% former.
\begin{lem}
  \label{lem:sum_j_bound}
  \begin{equation}
    \sum_{i = j+1}^\infty i \, Q_i c_1^i
    \leq
    C j\, Q_{j+1} c_1^{j+1}
    \quad
    \text{ for all } j \geq 1,
  \end{equation}
  where $C$ can be taken to be
  \begin{equation}
    C = 3 \frac{z_s^2}{(z_s-c_1)^2}.
  \end{equation}
\end{lem}

\begin{proof}
  Using
  % lemma \ref{lem:sum},
  the hypothesis that $Q_i z_s^i$ is decreasing in $i$, and calling $r
  := c_1/z_s$,
  \begin{multline}
    \sum_{i = j+1}^\infty i \, Q_i c_1^i \leq Q_{j+1} z_s^{j+1}
    \sum_{i = j+1}^\infty i \paren{\frac{c_1}{z_s}}^i
    \\
    = Q_{j+1} z_s^{j+1} \paren{\frac{c_1}{z_s}}^{j+1}
    \sum_{i' = 0}^\infty (i'+j+1) \paren{\frac{c_1}{z_s}}^{i'}
    \\
    = Q_{j+1} c_1^{j+1}
    \paren{ \frac{r}{(1-r)^2} + (j+1) \frac{1}{1-r} }
    \\
    \leq 3 \frac{1}{(1-r)^2} j\, Q_{j+1} c_1^{j+1} = C j\, Q_{j+1}
    c_1^{j+1},
  \end{multline}
  where the last inequality is obtained by observing that both
  $\frac{r}{(1-r)^2}$ and $\frac{1}{1-r}$ are smaller than
  $\frac{1}{(1-r)^2}$, and that $j+2 \leq 3 j$ for all $j \geq 1$.
\end{proof}

% The letters $C, C', C'',\dots$, $C_1, C_2, \dots$ will be used to
% denote constants which depend on the quantities allowed in the
% statement of proposition \ref{prp:mass_difference_bound}, and in the
% way specified there (this is, depending increasingly on $c_1/z_s$).
% For short, we will refer to these as ``allowed constants''. For
% convenience, they may refer to different constants in the same proof,
% but this will be explicitly indicated.

\begin{proof}[Proof of proposition \ref{prp:mass_difference_bound}]
  Call
  \begin{equation}
    u_i := \frac{c_i}{Q_i c_1^i}.
  \end{equation}
  With this notation, $D$ can be rewritten as
  \begin{equation}
    \label{eq:D_with_ui}
    D
    =
    \sum_{i = 1}^\infty
    a_i Q_i c_1^{i+1}
    \paren{
      u_i - u_{i+1}
    }
    \paren{
      \log u_i - \log u_{i+1}
    }.
  \end{equation}
  Noting that $u_1 = 1$ and using lemma \ref{lem:sum_j_bound} we can
  write
  \begin{multline}
    \sum_{i \geq 1} i\, c_i - \sum_{i \geq 1} i\,Q_i c_1^i
    =
    \sum_{i \geq 1} i\, Q_i c_1^i (u_i - 1)
    \\
    = \sum_{i \geq 1} i\, Q_i c_1^i \sum_{j=1}^{i-1}(u_{j+1} - u_j) =
    \sum_{j = 1}^\infty (u_{j+1} - u_j) \sum_{i = j+1}^\infty i \, Q_i
    c_1^i
    \\
    \leq \sum_{\underset{u_{j+1} \geq u_j}{j \geq 1}} (u_{j+1} - u_j)
    \sum_{i = j+1}^\infty i \, Q_i c_1^i \leq C_1
    \sum_{\underset{u_{j+1} \geq u_j}{j \geq 1}} j\, Q_{j+1} c_1^{j+1}
    (u_{j+1} - u_j)
    \\
    \leq \frac{C_1}{\sqrt{z_s}}
    \paren{ \sum_{\underset{u_{j+1} \geq u_j}{j \geq 1}} a_j Q_j c_1^{j+1}
      \frac{(u_{j+1} - u_j)^2}{u_{j+1}} }^{1/2}
    \paren{ \sum_{j \geq 1} \frac{j^2}{a_j} \, Q_{j+1} c_1^{j+1}
      u_{j+1} }^{1/2},
  \end{multline}
  where $C_1$ is the constant in lemma \ref{lem:sum_j_bound}. Here we
  have used the Cauchy-Schwarz inequality and the fact that $z_s
  Q_{j+1} \leq Q_j$ for all $j$ (hypothesis \ref{hyp:Qkj_regular}).
  Now, with
  \begin{equation}
    \label{eq:square-log-bound}
    \frac{(x-y)^2}{\max\{x,y\}} \leq (x-y)(\log x - \log y)
    \quad \text{ for all } x,y > 0,
  \end{equation}
  one sees from equation \eqref{eq:D_with_ui} that the first
  parenthesis is less than $D$. For the second one, use hypothesis
  \ref{hyp:coag_small_particles} to write
  \begin{equation}
    \sum_{j \geq 1} \frac{j^2}{a_j} \, Q_{j+1} c_1^{j+1} u_{j+1}
    \leq
    \frac{1}{K_1} \sum_{j \geq 1} (j+1)^{2-\lambda} c_{j+1}
    \leq
    \frac{1}{K_1} M_{2-\lambda}.
  \end{equation}
  We finally obtain that
  \begin{equation}
    \sum_{i \geq 1} i\, c_i - \sum_{i \geq 1} i\,Q_i c_1^i
    \leq
    C \sqrt{D} \sqrt{M_{2-\lambda}},
  \end{equation}
  with $C := \frac{C_1}{\sqrt{z_s}} \frac{1}{\sqrt{K_1}}$ (we recall
  that $C_1$ is the constant in lemma \ref{lem:sum_j_bound} and that
  $K_1$ is defined in hypothesis \ref{hyp:coag_small_particles}; note
  that $C_1$ is a constant with the dependence described in
  proposition \ref{prp:mass_difference_bound}). This proves the
  proposition.
\end{proof}

% \begin{lem}
%   \label{lem:sum}
%   For $r > 1$,
%   \begin{equation}
%     \sum_{i=0}^\infty i r^i = \frac{r}{(1-r)^2}.
%   \end{equation}
% \end{lem}

% \begin{proof}
%   We know that
%   \begin{equation}
%     \label{eq:sum_0}
%     \sum_{i=0}^\infty r^i = \frac{1}{1-r},
%   \end{equation}
%   and usual results on the derivation of power series allow us to
%   deduce that
%   \begin{equation}
%     \frac{d}{dr} \sum_{i=0}^\infty r^i
%     =
%     \sum_{i=1}^\infty i\, r^{i-1}
%     =
%     \sum_{i=0}^\infty (i+1)\, r^i
%     =
%     \sum_{i=0}^\infty i\, r^i
%     + \frac{1}{1-r}.
%   \end{equation}
%   Together with \eqref{eq:sum_0} this shows that
%   \begin{equation}
%     \sum_{i=0}^\infty i\, r^i
%     =
%     \frac{1}{(1-r)^2}
%     - \frac{1}{1-r}
%     =
%     \frac{r}{(1-r)^2}.
%   \end{equation}
% \end{proof}

\section{Strong convergence to equilibrium}
\label{sec:strong-conv-equil}

Let $c = \{c_i\}_{i \geq 1}$ be the solution to the DCF equations
\eqref{eq:DCF} on $[0, +\infty)$ with initial data $c^0 = \{c_i^0\}_{i
  \geq 1}$ given by proposition \ref{prp:existence_discrete} under
hypotheses \ref{hyp:growth}--\ref{hyp:moment_1+l_bounded}. Below we
will always denote by $\rho$ the mass of the solution $c$, which is
constant:
\begin{equation}
  \rho := \sum_{i \geq 1} i\, c_i(t)
  \quad \text{ for any } t \geq 0.
\end{equation}
Of course, if $\rho = 0$ then the solution itself is constantly 0 and
is uninteresting, so we will assume that $\rho > 0$. Our main result,
theorem \ref{thm:main-intro}, is stated more precisely as follows:
\begin{thm}
  \label{thm:main}
  Assume the hypotheses
  \ref{hyp:growth}--\ref{hyp:moment_1+l_bounded}. If $\rho > \rho_s$,
  then
  \begin{equation}
    c_i(t) \to Q_i z_s^i
    \quad \text{ for all } i \geq 1,
  \end{equation}
  while if $\rho \leq \rho_s$, then $c$ converges strongly to the only
  equilibrium with mass $\rho$:
  \begin{equation}
    \sum_{i \geq 1} i \abs{c_i(t) - Q_i z^i} \to 0
    \quad \text{ when } t \to +\infty
  \end{equation}
  for the only $z \geq 0$ such that
  \begin{equation}
    \rho = \sum_{i \geq 1} i\, Q_i z^i.
  \end{equation}
\end{thm}
By well-known arguments (see for example \cite{BCP86, CdC94,
  MR2186002}) this theorem follows from proposition
\ref{prp:weak-star-convergence} if we can show that whenever a
solution converges weak-$*$ to an equilibrium of mass strictly below
$\rho_s$, then the convergence must also be strong; the latter result
will be proved below in proposition \ref{prp:strong_convergence}. The
reason that this is enough is the following: by proposition
\ref{prp:weak-star-convergence}, we know that \emph{every} solution
must converge, at least weak-$*$, to some equilibrium with its same
mass or less. Then, if one has proposition
\ref{prp:strong_convergence}, one obviously has theorem \ref{thm:main}
for any solution with mass $\rho < \rho_s$. For a solution with mass
$\rho = \rho_s$, the weak-$*$ limit must be the equilibrium with mass
$\rho_s$, as any other limit with mass strictly less than $\rho_s$
implies strong convergence by proposition
\ref{prp:strong_convergence}, which is absurd (a strong limit must
have the same mass as the solution which converges to it). By lemma
\ref{lem:weak-strong-convergence}, the convergence to the equilibrium
with mass $\rho_s$ must be strong, as both masses coincide. This
proves theorem \ref{thm:main} for a solution with mass $\rho =
\rho_s$. Finally, for a solution with mass $\rho > \rho_s$, the same
argument shows that its only possible weak-$*$ limit is the
equilibrium with mass $\rho_s$, which completes the statement of
theorem \ref{thm:main}. For further detail on this, the reader can
look at the references mentioned above (\cite{BCP86, CdC94,
  MR2186002}).

\begin{prp}
  \label{prp:strong_convergence}
  If the solution $c$ converges weak-$*$ to an equilibrium $\{Q_i
  z^i\}$ with $0 \leq z < z_s$, then $\rho < \rho_s$, and $z$ is the
  only number such that
  \begin{equation}
    \rho = \sum_{i \geq 1} i\, Q_i z^i := \rho_z
  \end{equation}
  and the convergence is strong, in the sense that
  \begin{equation}
    \sum_{i \geq 1} i \abs{c_i(t) - Q_i z^i} \to 0
    \quad \text{ when } t \to +\infty.
  \end{equation}
\end{prp}

The bound in the previous section will be the fundamental tool to
prove the above proposition. In addition, we will need the following
two lemmas: the first one is a simple inequality which has been often
used in this context (see \cite[Appendix D]{C06} for a discussion of
this inequality and related ones), and which we prove for
completeness. It will be used to prove a bound on the increase of the
moment of order $2-\lambda$ of a solution, given in lemma
\ref{lem:moment_bound}.

\begin{lem}
  \label{lem:power-inequality}
  For $0 \leq \lambda \leq 1$ and $1 \leq k \leq 2-\lambda$ there is a
  constant $C_{k,\lambda} \geq 0$ such that
  \begin{equation}
    (x^\lambda + y^\lambda) ((x+y)^k - x^k - y^k)
    \leq
    C_{k,\lambda} (xy)^{\frac{\lambda + k}{2}}
    \quad
    \text{ for all } x,y \geq 0.
  \end{equation}
\end{lem}

\begin{proof}
  If any of $x$ or $y$ is zero, the inequality is trivial, so take $x,
  y > 0$. By symmetry, it is clearly enough to prove it when $x \leq
  y$. To do this, call $r := x/y$, so that $0 < r \leq 1$. We have
  \begin{multline}
    (1 + r^\lambda)((1+r)^k - 1 - r^k)
    \leq
    (1 + 2^\lambda) (k r (1+r)^{k-1} - r^k)
    \\
    \leq
    (1 + 2^\lambda) 2^{k-1} k r
    \leq
    (1 + 2^\lambda) 2^{k-1} k r^{\frac{\lambda + k}{2}}.
  \end{multline}
  In the first inequality we have used the mean value theorem and $k
  \geq 0$; in the second one we have left out the negative term and
  used that $k \geq 1$ and $r \leq 1$; and for the third one we have
  used that $1 \geq (\lambda + k)/2$ and $0 < r \leq 1$. Now,
  multiplying the beginning and end of the previous inequality by
  $y^{\frac{\lambda + k}{2}}$ and recalling the definition of $r$
  gives the inequality of the lemma.
\end{proof}

\begin{rem}
  Note that the inequality is also true, but of no value, for $k < 1$,
  as then the part on the left is negative and that on the right is
  positive. Also, note that in the previous lemma the constant can be
  chosen to be independent of $k, \lambda$.
\end{rem}

\begin{lem}
  \label{lem:moment_bound}
  Under the hypotheses of proposition \ref{prp:strong_convergence},
  there is some constant $C > 0$ which depends only on
  $M_{2-\lambda}(0)$, $\rho$, and the constant $K$ in hypothesis
  \ref{hyp:growth} such that
  \begin{equation}
    M_{2 -\lambda}(t) := \sum_{i=1}^\infty i^{2-\lambda} c_i(t)
    \leq C\, (1 + t)
    \quad \text{ for all } t \geq 0.
  \end{equation}
\end{lem}

\begin{proof}
  Take $c^N$ to be the solution of the finite system of size $N$ from
  the beginning of section \ref{sec:existence-solutions} with initial
  data $\{c^0_i\}_{i \leq N}$, and set $c^N_i := 0$ for $i > N$. We
  will prove the estimate for any such solution and a constant $C$
  depending only on the quantities in the lemma (and hence independent
  of $N$), and then a usual argument \cite{BC90, MR2186002} allows us
  to pass to the limit and get the same bound for the complete
  solution $c$. In fact, we will denote $c^N$ as $c$ to simplify the
  notation. Using a well-known identity giving the time derivative of
  moments of the solution $c$ \cite{MR2186002} we have,
  \begin{multline}
    \frac{d}{dt} M_{2-\lambda}(t)
    \leq
    \frac{1}{2} \sum_{i,j=1}^\infty a_{i,j} c_i(t) c_j(t) ((i+j)^{2-\lambda}
    - i^{2-\lambda} - j^{2-\lambda})
    \\
    \leq
    K C'
    \sum_{i,j=1}^\infty c_i(t) c_j(t) ij
    =
    K C'
    \rho^2,
  \end{multline}
  where $K$ is the constant in hypothesis \ref{hyp:growth} and $C'$ is
  the one in lemma \ref{lem:power-inequality} for $k = 2-\lambda$.
  Then, for all $t \geq t_0$,
  \begin{equation}
    M_{2-\lambda}(t)
    \leq
    M_{2-\lambda}(0)
    + K C' \rho^2 t
  \end{equation}
  which proves the lemma with, for example, $C := M_{2-\lambda}(0) +
  K C' \rho^2$.
\end{proof}

Now, let us prove proposition \ref{prp:strong_convergence}:

\begin{proof}[Proof of proposition \ref{prp:strong_convergence}]
  It is enough to prove the first statement (that $\rho = \rho_z$, the
  mass of the equilibrium to which the solution converges weak-$*$),
  as then the strong convergence follows from lemma
  \ref{lem:weak-strong-convergence}.  Note that we already know that
  $\rho \geq \rho_z$ thanks to proposition
  \ref{prp:weak-star-convergence} --- only loss of mass, not gain, can
  take place in the large time limit --- so we only need to prove that
  $\rho \leq \rho_z$.

  As the solution $c$ converges weak-$*$ to $\{Q_i z^i\}$, we know
  that $c_1 \to z < z_s$ and after some time $t_0 > 0$ it holds that
  \begin{equation}
    c_1(t) \leq \frac{z+z_s}{2} < z_s
    \quad \text{ for all } t \geq t_0.
  \end{equation}
  Then, calling $\rho_1(t) := \sum_{i \geq 1} i\, Q_i c_1^i(t)$ and
  applying proposition \ref{prp:mass_difference_bound} to $c(t)$ for
  $t \geq t_0$ we have, for some fixed constants $C_1$, $C_2$,
  \begin{equation}
    \label{eq:mass_difference_bound}
    \rho - \rho_1(t)
    \leq
    C_1 \sqrt{D} \sqrt{M_{2-\lambda}(t)}
    \leq
    C_2 \sqrt{D} \sqrt{1 + t}
    \quad \text{ for } t \geq t_0,
  \end{equation}
  thanks to lemma \ref{lem:moment_bound}. Now we note that
  \begin{equation}
    \label{eq:limit_rho_1}
    \lim_{t \to +\infty} \rho_1(t)
    = \rho_z = \sum_{i \geq 1} i Q_i z^i,
  \end{equation}
  which is a consequence of the continuity in $z$ of the above power
  series (which has radius of convergence $z_s$) and the fact that
  $c_1(t) \to z$ as $t \to \infty$.

  We would like to obtain a lower bound for $D$ from equation
  \eqref{eq:mass_difference_bound}, but this can only be done when the
  left hand side is positive. Let us see that we can suppose this to
  hold after a certain time $t_1$: otherwise there is a sequence $t_n
  \to \infty$ such that $\rho - \rho_1(t_n) \leq 0$, or $\rho \leq
  \rho_1(t_n) \to \rho_z$, so $\rho \leq \rho_z$ and the statement is
  proved. So we can assume that there is a time $t_1 \geq t_0$ such
  that $\rho > \rho_1(t)$ for all $t \geq t_1$.

  Then, for all $t \geq t_1$, equation
  \eqref{eq:mass_difference_bound} implies that
  \begin{equation}
    D(t) \geq C_3 \frac{(\rho - \rho_1(t))^2}{1 + t}
    \quad \text{ for } t \geq t_1,
  \end{equation}
  for $C_3 := 1 / C_2^2$. Now, if $V$ represents the free energy of
  the solution $c$, we know that for $t \geq t_1$
  \begin{align}
    V(t)
    & = V(t_1) -\int_{t_1}^t D_{CF}(s) \,ds
    \\
    & \leq V(t_1) - \int_{t_1}^t D(s) \,ds
    \\
    & \leq V(t_1) - C_3 \int_{t_1}^t
    \frac{(\rho - \rho_1(s))^2}{1 + s} \,ds.
  \end{align}
  As $V$ is bounded below for all times, we see the right hand side
  must be bounded for all times $t \geq t_1$; hence, knowing from
  \eqref{eq:limit_rho_1} that $\rho_1(t)$ has a limit as $t \to
  \infty$, this proves that its limit is $\rho$. On the other hand,
  its limit is $\rho_z$ according to equation \eqref{eq:limit_rho_1},
  so it must be $\rho = \rho_z$, which finishes the proof.
\end{proof}

\section{Relative energy estimate}
\label{sec:relat-energy-estim}

Take a nonnegative sequence $c = \{c_i\}_{i \geq 1}$ with $0 < c_1 <
z_s$. We are interested in estimating the relative energy $F_{c_1}$ of
$c$ to $\{Q_i c_1^i\}_{i \geq 1}$, a strategy also used in
\cite{JN03}. For brevity, we denote $F \equiv F_{c_1}(c)$ and write
\begin{equation}
  u_i := \frac{c_i}{Q_i c_1^i},
\end{equation}
so that $F$ can be rewritten as
\begin{align}
  F
  & := \sum_{i=1}^\infty
  c_i  \paren{
    \log \frac{c_i}{Q_i c_1^i}
    - 1
  }
  + \sum_{i=1}^\infty Q_i c_1^i\\
  & = \sum_{i=1}^\infty
  Q_i c_1^i \paren{
    u_i \log u_i - u_i
    + 1
  }
  \\
  & =
  \sum_{i=1}^\infty  Q_i c_1^i f(u_i),
\end{align}
where
\begin{equation}
  f(x) := x \log x - x + 1
  \quad \text{ for } x > 0.
\end{equation}
% and note that $f'(x) = \log x$ for $x > 0$.
Note that $F$ is finite if $0 < c_1 < z_s$.

With the same notation $D = D(c)$ can be rewritten as in equation
\eqref{eq:D_with_ui}, which we recall here:
\begin{equation}
  \label{eq:D_with_ui-2}
  D
  =
  \sum_{i = 1}^\infty
  a_i Q_i c_1^{i+1}
  \paren{
    u_i - u_{i+1}
  }
  \paren{
    \log u_i - \log u_{i+1}
  }.
\end{equation}

  % \begin{equation}
%   D_{CF} =
%   \frac{1}{2}
%   \sum_{i,j= 1}^\infty
%   a_{i,j} Q_i Q_j c_1^{i+j}
%   \paren{
%     u_i u_j - u_{i+j}
%   }
%   \paren{
%     \log (u_i u_j) - \log u_{i+j}
%   }.
% \end{equation}

% % 
% Below we will use the free energy dissipation rate which appears in
% the Becker-Döring equations:
% \begin{equation}
% %  \label{eq:BD_dissipation}
%   D
%   :=
%   \sum_{i = 1}^\infty
%   a_i Q_i c_1^{i+1}
%   \paren{
%     u_i - u_{i+1}
%   }
%   \paren{
%     \log u_i - \log u_{i+1}
%   }
% \end{equation}
% where
% \begin{align}
%   &a_1 := \frac{1}{2} a_{1,1},\\
%   &a_i := a_{i,1} \quad \text{ for } i > 1,
% \end{align}
% and we set $D = +\infty$ whenever $c_i = 0$ for some $i \geq 1$, as we
% did for $D_{CF}$. Note that $0 \leq D \leq D_{CF}$, as every term in
% $D$ already appears in $D_{CF}$, taking into account the symmetry of
% $a_{i,j}$.

In this section we show the following result:
\begin{prp}
  \label{prp:F_bound}
  Assume hypotheses \ref{hyp:growth}--\ref{hyp:moment_1+l_bounded},
  and let $c = \{c_i\}_{i \geq 1}$ be a strictly positive sequence
  with $0 < c_1 < z_s$ and with $M_{2-\lambda} := \sum_{i \geq 1}
  i^{2-\lambda} c_i < +\infty$. Then there is some constant $C \geq 0$
  that depends only on the coefficients $a_{i,j}$, $b_{i,j}$ ($i,j
  \geq 1$), on $\rho$ and continuously on $c_1$ such that
  \begin{equation}
    F
    \leq
    C \max\{\sqrt{D}\sqrt{M_{2-\lambda}}, D\}
  \end{equation}
  where $D = D(c)$ is the Becker-Döring free energy dissipation term
  defined in \eqref{eq:BD_dissipation}.
%   and
%   \begin{equation}
%     \label{eq:def_G}
%     G :=
%     \paren{1 + \log \frac{z_s}{c_1}}
%     (1 + M_{2-\lambda}).
%   \end{equation}
\end{prp}

\begin{rem}
  \label{rem:continuous-dependence-c1}
  The constant $C$ in the previous proposition may become infinite as
  $c_1$ approaches 0 or $z_s$; however, we specify that it depends
  \emph{continuously} on $c_1$ so that, if one knows that $\epsilon <
  c_1 < z_s - \epsilon$ for some $\epsilon > 0$, then the constant may
  be taken to depend on $\epsilon$ and not on $c_1$. This will be used
  in the proof of proposition \ref{prp:rate-conv-equil}.
\end{rem}

\begin{rem}
  The dependence on $c_1$ of the above inequality may be of interest;
  for example, if one wants to use it to prove theorem \ref{thm:main}
  (instead of the inequality in proposition
  \ref{prp:mass_difference_bound}), one needs some control on the
  constant as $c_1 \to 0$ in order to rule out the possibility that
  solutions converge weakly to the equilibrium with mass 0 (i.e., $c_i
  \equiv 0$ for all $i$). We have not explicitly stated this
  dependence for simplicity (as it is not used, makes the proof
  somewhat more cumbersome, and after theorem \ref{thm:main} we know
  that $c_1$ is greater than some positive constant after a certain
  time anyway), but the reader can check from the constants in the
  proof that the growth of $C$ as $c_1 \to 0$ is controlled by
  $\abs{\log c_1}$.
\end{rem}

Let us prove the above inequality. Of course, the inequality is
nontrivial only when $D < +\infty$, so we assume that $D$ is finite.
The case $D = 0$ is also trivial, for if $D$ vanishes, $c$ must be a
nonzero equilibrium, and then $F = 0$; hence, we will also assume that
$D > 0$.

In the course of the present proof the letters $C, C_1, C_2, \dots$
will always be used to denote numbers which depend on the quantities
allowed in the statement of proposition \ref{prp:F_bound}, and in the
way specified there. For short, we will frequently refer to these as
``allowed constants''.

Take any integer $N \geq 1$ and split the sum in $F$ as
\begin{equation}
  F = \sum_{i \leq N}  Q_i c_1^i f(u_i)
  + \sum_{i > N}  Q_i c_1^i f(u_i)
  =: F_1 + F_2.
\end{equation}

\noindent
\textbf{First step: estimate for $F_1$.} As $u_1 = 1$ and $f(1) = 0$,
\begin{equation}
  f(u_i) = \sum_{j=1}^{i-1} (f(u_{j+1}) - f(u_j))
  \quad \text{ for } i \geq 1.
\end{equation}
Note that the sum is empty for $i=1$.
With this,
\begin{multline}
  \label{eq:F_1}
  F_1 = \sum_{i=1}^N  Q_i c_1^i f(u_i)
  =
  \sum_{i=1}^N \sum_{j=1}^{i-1} Q_i c_1^i (f(u_{j+1}) - f(u_j))
  \\
  = \sum_{j=1}^{N-1} (f(u_{j+1}) - f(u_j))
  \sum_{i=j+1}^N Q_i c_1^i
  \\
  \leq
  C_1 \sum_{\underset{f(u_{j+1}) \geq f(u_j)}{j=1}}^N
  Q_{j+1} c_1^{j+1} (f(u_{j+1}) - f(u_j)),
\end{multline}
where the last inequality, for some allowed constant $C_1$, is
obtained in a very similar way to that in lemma \ref{lem:sum_j_bound}.

% \begin{equation}
%   T_j
%   := \sum_{i=j+1}^N Q_i c_1^i
%   \quad \text{ for } 1 \leq j < N.
% \end{equation}
% %
% Let us estimate $T_j$:

% \begin{lem}
%   There is an allowed constant $C \geq 0$ such that
%   \begin{equation}
%     T_j \leq C Q_{j+1} c_1^{j+1}
%     \quad \text{ for all } j \geq 1.
%   \end{equation}
% \end{lem}

% \begin{proof}
%   The proof is based on the fact that the sum defining $T_j$ behaves
%   as the sum of $(c_1/z_s)^i$, which can be calculated. Owing to
%   hypothesis \ref{hyp:Qkj_regular} we have, for $j > M$,
%   \begin{multline}
%     T_j
%     \leq
%     \sum_{i=j+1}^\infty Q_i c_1^i
%     \leq
%     \sum_{i=j+1}^\infty \paren{\frac{c_1}{z_s}}^i Q_i z_s^i
%     \leq
%     Q_{j+1} z_s^{j+1}
%     \sum_{i=j+1}^\infty \paren{\frac{c_1}{z_s}}^i
%     \\
%     =
%     Q_{j+1} z_s^{j+1} \paren{\frac{c_1}{z_s}}^{j+1}
%     \frac{z_s}{z_s - c_1}
%     =
%     Q_{j+1} c_1^{j+1} \frac{z_s}{z_s - c_1},
%   \end{multline}
%   which proves the inequality for $j > M$, as $z_s/(z_s - c_1)$
%   depends increasingly on $c_1$ and is an allowed constant. For any
%   particular $j \leq M$ the statement is also true, as
%   \begin{equation}
%     \frac{T_j}{Q_{j+1} c_1^{j+1}}
%     = \frac{1}{Q_{j+1}} \sum_{i \geq j+1} Q_i c_1^{i-j-1},
%   \end{equation}
%   which is a convergent series, as $\lim_{i \to \infty} Q_i^{1/i} =
%   1/z_s$. In addition, observe that the value of the sum depends
%   increasingly on $c_1$ and is an allowed constant. Hence, the
%   inequality is also true for all $j \leq M$, as this involves only a
%   finite number of values of $j$, and this finishes the proof.
% \end{proof}

\begin{lem}
  For $x, y > 0$ it holds that
  \begin{equation}
    f(x) - f(y)
    \leq (x-y) (\log x - \log y)
    + (x-y) \log \max\{x,y\}.
  \end{equation}
\end{lem}

\begin{proof}
  Regardless of the sign of $x-y$, the mean value theorem shows that
  \begin{multline}
    f(x) - f(y)
    \leq
    (x-y) \log x
    \\
    = (x-y) (\log x - \log y) + (x-y) \log y
    \\
    \leq
    (x-y) (\log x - \log y) + (x-y) \log \max\{x,y\}.
  \end{multline}
  Again, notice that the last step holds both when $x \leq y$ and $y
  \leq x$.
\end{proof}

% 
% Also, when $u_{j+1} \geq u_j$,
% \begin{equation}
%   f(u_{j+1}) - f(u_j) \leq (u_{j+1} - u_j) \log u_{j+1}.
% \end{equation}
%

With the previous lemma we can continue from \eqref{eq:F_1}.  Denoting
$w_j := \max\{u_j, u_{j+1}\}$,
\begin{multline}
  \label{eq:F_1_Schwartz}
  F_1
  \leq C_1 \sum_{\underset{f(u_{j+1}) \geq f(u_j)}{j=1}}^N
  (f(u_{j+1}) - f(u_j)) Q_{j+1} c_1^{j+1}
  \\
  \leq
  C_1
  \sum_{j \leq N}
  Q_{j+1} c_1^{j+1}
  (u_{j+1} - u_j) (\log u_{j+1} - \log u_j)
  \\
  +
  C_1
  \sum_{j \leq N}
  Q_{j+1} c_1^{j+1}
  \abs{u_{j+1} - u_j} \abs{\log w_j}
  =:
  T_1 + T_2.
\end{multline}
For the first term, $T_1$, we can use once more that $z_s Q_{i+1} \leq
Q_i$ (hypothesis \ref{hyp:Qkj_regular}), the lower bound on $a_j$ from
hypothesis \ref{hyp:coag_small_particles} and the expression of $D$ in
eq. \eqref{eq:D_with_ui-2} to see that
\begin{equation}
  \label{eq:bound_T_1}
  T_1 \leq 2 \frac{C_1}{K_1 z_s} D =: C_2 D,
\end{equation}
where $K_1$ is the constant in hypothesis
\ref{hyp:coag_small_particles} (and the factor of 2 appears because of
the definition of $a_1$ in eq. \eqref{eq:def-a_1}). For the second
term in \eqref{eq:F_1_Schwartz}, $T_2$, the Cauchy-Schwarz inequality
gives
\begin{equation}
  T_2
  \leq
  C_3
  \paren{
    \sum_{j \leq N}
    a_j Q_{j} c_1^{j+1}
    \frac{(u_{j+1} - u_j)^2}{ w_j }
  }^{1/2}
  \paren{
    \sum_{j = 1}^N
    \frac{1}{a_j}
    Q_{j+1} c_1^{j+1}
    w_j (\log w_j)^2
  }^{1/2},
\end{equation}
where $C_3 := \frac{C_1}{\sqrt{z_s}}$, again using that $z_s Q_{i+1}
\leq Q_i$ (hypothesis \ref{hyp:Qkj_regular}). By inequality
\eqref{eq:square-log-bound} and eq. \eqref{eq:D_with_ui-2}, the first
term inside parentheses is less than $D$, so
\begin{equation}
  \label{eq:bound_T_2}
  T_2
  \leq
  C_3
  \sqrt{D}
  \paren{
    \sum_{j = 1}^N
    \frac{1}{a_j}
    Q_{j+1} c_1^{j+1}
    w_j (\log w_j)^2
  }^{1/2}.
\end{equation}
Now let us use the following result to compare $w_j (\log w_j)^2$ with
$f(w_j)$:

\begin{lem}
  \label{lem:xlogx_inequality}
  It holds that
  \begin{equation}
    x(\log x)^2 \leq 4 (x \log x - x + 1) \max\{1, \log x\}
    \quad \text{ for } x > 0.
  \end{equation}
\end{lem}

\begin{proof}
  Call $g(x) := x(\log x)^2$ and $f(x) := (x \log x - x + 1)$ as
  before. Then, $f(1) = f'(1) = g(1) = g'(1) = 0$, $f''(x) = 1/x$ and
  \begin{equation}
    g''(x) = 2 \frac{\log x}{x} + 2 \frac{1}{x}
    \leq 4 f''(x)
    \quad \text{ for } 0 < x \leq e,
  \end{equation}
  so by integrating one gets $g(x) \leq 4 f(x)$ for $0 < x \leq e$ and
  we have proved the inequality in this range.

  Now, for $x \geq e$, we have $\log x \geq 1$ and the inequality is
  equivalent to showing that
  \begin{equation}
    3 x \log x  - 4 (x-1) \geq 0
    \quad \text{ for } x \geq e,
  \end{equation}
  but the derivative of this function is $3 \log x -1$, which is
  clearly positive for $x \geq e$; hence, the function itself is
  greater than its value at $x = e$, which is $3 e - 4 (e-1) = 4 - e >
  0$. This finishes the proof.
%  
%   Now, for $x \geq e$, notice that the function $x \mapsto \frac{x
%     \log x}{x-1}$ is increasing, so for $x$ in this range, $x - 1 \leq
%   \frac{e-1}{e} x \log x$ and
%   \begin{equation}
%     f(x) \log x
%     = \log x (x \log x - x + 1)
%     \geq \frac{1}{e} \log x (x \log x)
%     \geq \frac{1}{e} g(x)
%     \quad (x \geq e),
%   \end{equation}
%   which finishes the proof.
\end{proof}

With the previous lemma,
\begin{equation}
  \frac{1}{a_j} w_j (\log w_j)^2
  \leq
  \frac{4}{a_j} f(w_j) \max\{1, \log w_j\}
  \leq 4 M_N f(w_j)
  \quad \text{ for } j \leq N,
\end{equation}
where $M_N$ is the maximum for $j \leq N$ of the expression
$\frac{1}{a_j} \max\{ 1, \log w_j \}$. Hence, continuing from
\eqref{eq:bound_T_2},
\begin{equation}
  \label{eq:bound_T_2-2}
  T_2
  \leq
  2\, C_3
  \sqrt{D}
  \sqrt{M_N}
  \paren{
    \sum_{j = 1}^N
    Q_{j+1} c_1^{j+1}
    f(w_j)
  }^{1/2}.
\end{equation}
Now note that
\begin{equation}
  f(w_j) \leq f(u_j) + f(u_{j+1})
\end{equation}
and again that $z_s Q_{j+1} \leq Q_j$ (hyp. \ref{hyp:Qkj_regular}) to
get
\begin{equation}
  \sum_{j = 1}^N
  Q_{j+1} c_1^{j+1}
  f(w_j)
  \leq
  \frac{c_1}{z_s}
  \sum_{j = 1}^{N}
  Q_{j} c_1^{j}
  f(u_j)
  +
  \sum_{j = 1}^{N}
  Q_{j+1} c_1^{j+1}
  f(u_{j+1})
  \leq
  C_4\, F,
\end{equation}
where $C_4$ can be taken to be $1 + c_1/z_s$, an allowed constant.
Hence, from \eqref{eq:bound_T_2-2},
\begin{equation}
  \label{eq:bound_T_2-3}
  T_2
  \leq
  C_5
  \sqrt{D}
  \sqrt{M_N}
  \sqrt{F},
\end{equation}
with $C_5 := 2\, C_3 \sqrt{C_4}$. Observe that, as $c_i \leq \rho/i$
for $i \geq 1$,
\begin{equation}
  \log u_i
  \leq \log \frac{\rho}{i Q_i c_1^i}
  = i \left(
    \log \frac{\rho^{1/i}}{i^{1/i} Q_i^{1/i} } + \log \frac{1}{c_1}
  \right)
  \leq
  C_7\, i
\end{equation}
for some allowed constant $C_7$. We have used that $Q_i^{1/i}$ is
bounded below by some constant thanks to hypothesis \ref{hyp:Qjk}, and
thus the term inside the parentheses is bounded above by some allowed
constant. Knowing that $a_i \geq K_1 i^\lambda$ (hyp.
\ref{hyp:coag_small_particles}),
\begin{equation}
  M_N \leq 2 \, C_7 \frac{1}{K_1} N^{1-\lambda} =: C_8 N^{1-\lambda},
\end{equation}
and from~\eqref{eq:bound_T_2-3},
\begin{equation}
  \label{eq:bound_T_2-4}
  T_2
  \leq
  C_9
  \sqrt{D}
  \sqrt{N^{1-\lambda}}
  \sqrt{F},
\end{equation}
with $C_9 := C_5 \sqrt{C_8}$. Now, putting together
\eqref{eq:F_1_Schwartz}, \eqref{eq:bound_T_1} and
\eqref{eq:bound_T_2-4} we have
\begin{align}
  \nonumber
  F_1
  & \leq
  C_2 D + C_9 \sqrt{N^{1-\lambda} D}
  \sqrt{F}
  \\
  \nonumber
  & \leq
  C_2 D
  +
  \frac{1}{2} F + \frac{C_9^2}{2} N^{1-\lambda} D
  \\
  \label{eq:F_1_bound}
  & \leq
  \frac{1}{2} F + C_{10}\, N^{1-\lambda} D,
\end{align}
with $C_{10} := C_2 + C_9^2 / 2$.

\noindent
\textbf{Second step: Estimate for $F_2$.} \emph{(In this step, the
  symbols $C_1, C_2, \dots$ are used again for convenience to denote
  allowed constants, but they have nothing to do with previous
  appearances of them)}. We have
\begin{align}
  \nonumber
  F_2
  & = \sum_{i > N} Q_i c_1^i u_i \log u_i
  - \sum_{i > N} c_i
  + \sum_{i > N} Q_i c_1^i
  \\
  \label{eq:F2_bound_prev}
  & \leq \sum_{i > N} Q_i c_1^i u_i \log u_i
  + \frac{1}{N} \sum_{i > N} i Q_i c_1^i
  \\
  \label{eq:F2_bound_prev-2}
  & \leq \sum_{i > N} Q_i c_1^i u_i \log u_i
  + \frac{1}{N} C_1,
\end{align}
where $C_1$ is $\sum_{i \geq 1} i Q_i c_1^i$, an allowed constant. For
the other term one has, writing $\Psi(x) := x \log x$ (a superadditive
function) and taking some constant $1 \geq C_2 > 0$ such that $Q_i^{1/i}
\geq C_2$ (which is possible by hyp. \ref{hyp:Qjk}),
\begin{align}
  \sum_{i > N} Q_i c_1^i u_i \log u_i
  & = \sum_{i > N} (Q_i c_1^i u_i) \log (Q_i c_1^i u_i)
  - \sum_{i > N} Q_i c_1^i u_i \log (Q_i c_1^i)
  \\
  & \leq \Psi\paren{ \sum_{i > N} c_ i }
  + \log \frac{C_2}{c_1} \sum_{i > N}  i c_i.
\end{align}
Now, take $N \geq \rho$, so that $\sum_{i > N} c_ i \leq \rho/N \leq
1$, which makes the first term negative. Then, calling $C_3 := \log
\frac{C_2}{c_1}$ and continuing from above,
\begin{equation}
  \sum_{i > N} Q_i c_1^i u_i \log u_i
  \leq
  \frac{C_3}{N^{1-\lambda}}
  \sum_{i > N}  i^{2-\lambda} c_i
  \leq
  \frac{C_3}{N^{1-\lambda}}
  M_{2-\lambda}.
\end{equation}
Together with \eqref{eq:F2_bound_prev-2} we obtain
\begin{equation}
  \label{eq:F_2-bound}
  F_2
  \leq
  \frac{C_3}{N^{1-\lambda}}
  M_{2-\lambda}
  +
  \frac{C_1}{N}
  \leq
  \frac{C_4}{N^{1-\lambda}} M_{2-\lambda},
\end{equation}
with $C_4 := C_1 + C_3$.

\noindent
\textbf{Third step: Estimate for $F$.} \emph{Again in this step,
  constants $C_1,C_2,\dots$ have nothing to do with previous ones
  unless explicitly noted}. With \eqref{eq:F_1_bound} and
\eqref{eq:F_2-bound} we have, for any $N \geq \rho$,
\begin{equation}
  F
  \leq
  \frac{1}{2} F
  + C_{10} N^{1-\lambda} D
  +
  \frac{C_4}{N^{1-\lambda}} M_{2-\lambda},
\end{equation}
where $C_{10}$ is the constant from eq. \eqref{eq:F_1_bound} and $C_4$
is that from eq. \eqref{eq:F_2-bound}. Hence, taking $C := \max\{2
  C_{10}, 2 C_4\}$,
\begin{equation}
  F
  \leq
  C \left(
    N^{1-\lambda} D
    +
    \frac{1}{N^{1-\lambda}} M_{2-\lambda}
  \right)
  \quad \text{ for all integers } N \geq \rho.
\end{equation}
% \begin{align}
%   & \leq
%   \frac{2(C + \log \frac{z_s}{c_1})}{N^{1-\lambda}}
%   (1 + M_{2-\lambda})
%   + 2 C_6 N^{1-\lambda} D
%   \\
%   & \leq 
%   \frac{C_7\, G}{N^{1-\lambda}}
%   + 2\, C_6 N^{1-\lambda} D
%   \\
%   & \leq 
%   \frac{C\, G}{N^{1-\lambda}}
%   + C N^{1-\lambda} D,
% \end{align}
Actually, it is clear that if we take $C_1 := 2 C$ one can write the
above for all real $N$ such that $R := N^{1-\lambda} \geq (\rho +
2)^{1-\lambda} =: C_2$ instead of only for the integers, just by
applying the previous inequality to the integer closest to $N$:
\begin{equation}
  \label{eq:F-bound-to-optimize}
  F
  \leq
  C_1 \left(
    R D
    +
    \frac{1}{R} M_{2-\lambda}
  \right)
  \quad \text{ for all real } R
  \text{ with } R \geq C_2.
\end{equation}
Let us choose $R$ in a way that gives a suitable inequality:
\begin{itemize}
\item If $\frac{\sqrt{M_{2-\lambda}}}{\sqrt{D}} \geq C_2$,
  then we take $R := \frac{\sqrt{M_{2-\lambda}}}{\sqrt{D}}$ and we
  obtain
  \begin{equation}
    \label{eq:F-ineq-1}
    F \leq C_1 \sqrt{D} \sqrt{M_{2-\lambda}}.
  \end{equation}
\item Otherwise, if $\frac{\sqrt{M_{2-\lambda}}}{\sqrt{D}} < C_2$
  then $M_{2-\lambda} \leq C_2^2 D$ and inequality
  \eqref{eq:F-bound-to-optimize} with $R :=
  C_2$ gives
  \begin{equation}
    \label{eq:F-ineq-2}
    F
    \leq C_1 C_2 D + \frac{C_1}{C_2} M_{2-\lambda}
    \leq C_1 C_2 D + \frac{C_1}{C_2} C_2^2 D
    = C_3 D,
  \end{equation}
  with $C_3 := 2 C_1 C_2.$
\end{itemize}
Equations \eqref{eq:F-ineq-1} and \eqref{eq:F-ineq-2} prove that, for
$C := \max\{C_1, C_3\}$,
\begin{equation}
  F
  \leq
  C \max\{\sqrt{D}\sqrt{M_{2-\lambda}}, D\},
\end{equation}
which proves the result.

\section{Rate of convergence to equilibrium}
\label{sec:rate-conv-equil}

With the previous results one can easily obtain the following rate of
convergence to equilibrium:

\begin{prp}
  \label{prp:rate-conv-equil}
  Let $c$ be a solution to the DCF equations given by proposition
  \ref{prp:existence_discrete} under hypotheses
  \ref{hyp:growth}--\ref{hyp:moment_1+l_bounded}. Suppose that the
  mass $\rho$ of the solution $c$ is strictly less than the critical
  mass $\rho_s$, and that $M_{2-\lambda}(0) := \sum_{i \geq 1}
  i^{2-\lambda} c_i(0) < +\infty$.  Then for some constant $C$
  depending only on the coefficients $a_{i,j}, b_{i,j}$, on
  $M_{2-\lambda}(0)$ and on the mass $\rho$,
  \begin{equation}
    F_z(t)
    \leq
    \min\left\{ F_z(0), \frac{C}{1 + \log(1+t)} \right\}
    \quad \text { for all } t > 0,
  \end{equation}
  where $z$ is such that
  \begin{equation}
    \sum_{i \geq 1} i Q_i z = \rho.
  \end{equation}
\end{prp}

\begin{rem}
  By inequality \eqref{eq:F_bounds_strong_convergence}, this implies
  that (for some other $C$ depending on the same quantities)
  \begin{equation}
    \sum_{i \geq 1} i \abs{ c_i - z^i Q_i }
    \leq \frac{C}{\sqrt{1 + \log t}}
    \quad \text { for all } t > 0.
  \end{equation}
\end{rem}

This rate is by no means expected to be optimal; in fact, one would
expect the solution to converge to equilibrium at a rate similar to
the one obtained by Jabin and Niethammer in \cite{JN03}, this is, a
convergence like $e^{-C t^{1/3}}$ for some constant $C > 0$. However,
further estimates on the solution (such as, for example, uniform
estimates on exponential moments) which are not readily available here
are essential in \cite{JN03} in order to show such convergence.

In order to prove proposition \ref{prp:rate-conv-equil} we will use
proposition \ref{prp:F_bound} and the following lemma from \cite[lemma
3.6]{JN03}, which is also applicable in our case:
\begin{lem}
  \label{lem:speed_big_c1}
  Assume hypotheses \ref{hyp:growth}--\ref{hyp:Qkj_regular}. Let $c =
  \{c_i\}_{i \geq 1}$ be a nonnegative sequence with mass $\rho <
  \rho_s$ and such that $c_1 \geq
  z_s - \frac{1}{4}(z_s - z)$, where $z$ is such that
  \begin{equation}
    \sum_{i \geq 1} i Q_i z^i = \rho,
  \end{equation}
  as usual. There exists a constant $C > 0$ which depends only on the
  $a_{i,j}, b_{i,j}$ and $\rho$ such that
  \begin{equation}
    D(c) \geq C.
  \end{equation}
\end{lem}

\begin{proof}[Proof of proposition \ref{prp:rate-conv-equil}]
  As before, $C, C_1, C_2, \dots$ are used to denote constants which
  depend on the quantities stated in the proposition, which will be
  called ``allowed constants''.
  
  First, note that $F_z$ is always finite under these conditions. In
  fact, one can see that $F_z(t)$ is bounded for all $t \geq 0$ by a
  constant $C_1$ which depends only on the coefficients and on $\rho$
  (equivalently, the free energy $V(t)$ is bounded by such a constant;
  as mentioned at the beginning of section \ref{sec:Lyapunov}, this
  result can be found in \cite[Lemma 4.2 and p. 680]{BCP86}, and can
  also be deduced from lemmas \ref{lem:moment_k+logc},
  \ref{lem:moment_k+logQ} and the expression of $F_z$ in eq.
  \eqref{eq:F_z-as-difference-of-free-energy}). So, with the H-theorem
  \ref{thm:H}, one has
  \begin{equation}
    \label{eq:F-unif-bounded}
    F_z(t) \leq F_z(0) \leq C_1
    \quad \text{ for all } t \geq 0.
  \end{equation}

  In fact, by the H-theorem we know that
  \begin{equation}
    \frac{d}{dt} F_z(t) = -D_{CF}(t) \leq -D(t),
  \end{equation}
  where $D_{CF}$ and $D$ are the dissipation rates defined in section
  \ref{sec:Lyapunov}. In order to use the inequality in proposition
  \ref{prp:F_bound} we note that after \cite[lemma 3.8]{JN03},
  whenever $0 < c_1 < z_s$ we have
  \begin{equation}
    \label{eq:Fz_leq_F}
    F_z(c) \leq F(c) < +\infty,
  \end{equation}
  where $F$ is just $F_{c_1}$ (the same used in proposition
  \ref{prp:F_bound}, and defined before it). So, we can use
  proposition \ref{prp:F_bound} to get a closed equation \emph{only}
  when $0 < c_1(t) < z_s$; in fact, if we want a bound which is
  independent of $c_1$, we need to use the inequality only for times
  $t$ for which $\epsilon < c_1(t) < z_s - \epsilon$ (see remark
  \ref{rem:continuous-dependence-c1}). Hence, we break the argument in
  three parts: when $c_1(t)$ is close enough to $z$, we use
  proposition \ref{prp:F_bound}; when $c_1(t)$ is above this region,
  we use the bound in lemma \ref{lem:speed_big_c1}, which controls the
  dissipation rate $D$ when $c_1$ is ``supercritical''; and for
  $c_1(t)$ below this region, we use the inequality in proposition
  \ref{prp:mass_difference_bound}, which is weaker than that in prop.
  \ref{prp:F_bound} but holds uniformly for small $c_1$. Let us do this:
  \begin{enumerate}
  \item At any time $t$ at which $c_1(t) \geq z_s - \frac{1}{4}(z_s -
    z)$, lemma \ref{lem:speed_big_c1} shows that for some allowed
    $C_2 > 0$,
    \begin{equation}
      D(t) \geq C_2.
    \end{equation}
    In order to use it below, note that by
    eq. \eqref{eq:F-unif-bounded} this can be bounded by
    \begin{equation}
      \label{eq:upper-region}
      D(t)
      \geq \frac{C_2}{C_1^2} \frac{F_z^2(t)}{1+t}
      \quad \text{ for all } t \geq 0.
    \end{equation}
  \item At any time such that $\frac{z}{2} \leq c_1(t) < z_s -
    \frac{1}{4}(z_s - z)$, proposition \ref{prp:F_bound}, and equation
    \eqref{eq:Fz_leq_F} show that
    \begin{equation}
      F_z(t) \leq F(t)
      \leq
      C_3 \max\{\sqrt{D(t)} \sqrt{M_{2-\lambda}(t)}, D(t)\},
    \end{equation}
    where $C_3$ is an allowed constant (which bounds the constant
    called $C$ in proposition \ref{prp:F_bound} for the $c_1$ under
    consideration). Then,
    \begin{equation}
      D(t)
      \geq
      \min\{
      \frac{1}{C_3^2} \frac{F_z^2(t)}{M_{2-\lambda}(t)},
      \frac{1}{C_3} F_z(t)
      \},
    \end{equation}
    and with lemma \ref{lem:moment_bound},
    \begin{equation}
      \label{eq:middle-region-prev}
      D(t)
      \geq
      \min\{
      \frac{1}{C_4 C_3^2} \frac{F_z^2(t)}{1+t},
      \frac{1}{C_3} F_z(t)
      \},
    \end{equation}
    where $C_4$ is the constant which appears in lemma
    \ref{lem:moment_bound} (called $C$ there). Now, using eq.
    \eqref{eq:F-unif-bounded},
    \begin{equation}
      \frac{F_z}{C_3}
      \geq \frac{F_z^2}{C_3 C_1}
      \geq \frac{F_z^2}{C_3 C_1 (1+t)},
    \end{equation}
    so from \eqref{eq:middle-region-prev} we have, for some allowed
    $C_5 > 0$,
    \begin{equation}
      \label{eq:middle-region}
      D(t)
      \geq
      C_5 \frac{F_z^2(t)}{1+t}.
    \end{equation}
  \item At any time such that $c_1(t) < \frac{z}{2}$, the
    inequality in proposition \ref{prp:mass_difference_bound} and
    again lemma \ref{lem:moment_bound} show that for some allowed
    constants $C_6$, $C_7$,
    \begin{equation}
      C_6 \leq
      \sum_{i \geq 1} i c_i(t) - \sum_{i \geq 1} i Q_i c_1(t)^i
      \leq
      C_7 \sqrt{D(t)} \sqrt{M_{2-\lambda}(t)},
    \end{equation}
    so, with $C_8 := (C_6/C_7)^2$,
    \begin{equation}
      \label{eq:lower-region}
      D(t)
      \geq C_8 \frac{1}{M_{2-\lambda}(t)}
      \geq \frac{C_8}{C_4 C_1^2} \frac{F_z^2(t)}{1+t},
    \end{equation}
    where $C_1$ appears in eq. \eqref{eq:F-unif-bounded} and $C_4$ is
    again the constant in lemma \ref{lem:moment_bound}.
  \end{enumerate}
  Hence, gathering eqs. \eqref{eq:upper-region},
  \eqref{eq:middle-region} and \eqref{eq:lower-region}, we know that
  there is an allowed constant $C_9$ such that
  \begin{equation}
    \frac{d}{dt} F_z(t)
    \leq - D(t)
    \leq - C_9 \frac{F_z^2(t)}{1+t}
    \quad \text{ for all } t \geq 0.
  \end{equation}
  Solving this differential inequality proves the proposition.
\end{proof}

\section{Appendix: Proof of the H-theorem}
\label{sec:H_theorem}

In this section we give the proof of the H-theorem \ref{thm:H}. The
usual strategy to prove this result is to calculate the time
derivative of an approximation to $V$ for which we know how to do it,
and then show that the limit behavior of these approximations imply
the H-theorem for $V$. We will follow this idea in a way similar to
the proof of \cite[Theorem 5.2]{CdC94}. On the way, we will make use
of some simple bounds stated in the following two lemmas:

\begin{lem}
  \label{lem:moment_k+logc}
  Take $m > k \in \RR$ with $m \geq 1$. For any nonnegative sequence
  $\{c_i\}$ there is a constant $C$ which depends only on $k$, $m$ and
  $M := \sum_{i=1}^\infty i^m c_i$ such that
  \begin{equation}
    \sum_{i=1}^\infty i^k c_i \abs{\log c_i}
    \leq C.
  \end{equation}
\end{lem}

\begin{proof}
  For any $0 < \epsilon < 1$ there is a constant $C_\epsilon \geq 0$
  such that $\abs{x \log x} \leq C_\epsilon (x^{1-\epsilon} +
  x^{1+\epsilon})$, so we have
  \begin{equation}
    \sum_{i=1}^\infty i^k c_i \abs{\log c_i}
    \leq
    C_\epsilon
    \paren{
      \sum_{i=1}^\infty i^k c_i^{1-\epsilon}
      +
      \sum_{i=1}^\infty i^k c_i^{1+\epsilon}
    }.
  \end{equation}
  As $c_i \leq M$ for all $i \geq 1$, the second sum is less than $M
  \sum_{i=1}^\infty i^k c_i \leq M^2$. For the first sum, using
  Hölder's inequality with exponents $p = 1/(1-\epsilon)$, $q =
  1/\epsilon$,
  \begin{equation}
    \sum_{i=1}^\infty i^k c_i^{1-\epsilon}
    \leq
    \paren{\sum_{i=1}^\infty i^m c_i}^{1-\epsilon}
    \paren{
      \sum_{i=1}^\infty
      i^{\frac{k - m(1-\epsilon)}{\epsilon}}
    }^\epsilon.
  \end{equation}
  As $k < m$, we can choose $\epsilon > 0$ small enough such that the
  exponent of $i$ inside the second sum is less than $-1$; with such
  an $\epsilon$, the sum is finite and the result is proved.
\end{proof}

\begin{lem}
  \label{lem:moment_k+logQ}
  Take a strictly positive sequence $\{Q_i\}$ such that $C_1 \geq
  Q_i^{1/i} \geq C_2$ for some $C_1 \geq C_2 > 0$ and all $i \geq 1$. Then
  for any nonnegative sequence $\{c_i\}$
  \begin{equation}
    \sum_{i=1}^\infty i^k c_i \abs{\log Q_i}
    \leq
    C
    \sum_{i=1}^\infty i^{k+1} c_i,
  \end{equation}
  with $C := \max\{\abs{\log C_1}, \abs{\log C_2}\}$.
\end{lem}

\begin{proof}
  One just writes $\abs{\log Q_i} = i \abs{\log Q_i^{1/i}}$ and use
  the bounds assumed in the lemma.
\end{proof}

Let us prove theorem \ref{thm:H}. First, note that the hypotheses that
$a_{i,1} > 0$ for $i \geq 1$ implies that $c_i(t)$ is strictly
positive for all $t > 0$ and all $i \geq 1$ \cite{C92, CdC94}, so that
$D_{CF}$ makes sense for all positive times. Note also that moments
which are finite at $t=0$ remain finite for all times \cite{BC90}; in
particular, the moments of order $1+\lambda$ and $1+\gamma$ are always
finite under our assumptions. Call
\begin{equation}
  V_N :=
  \sum_{i=1}^N
  c_i  \paren{
    \log \frac{c_i}{Q_i}
    - 1
  }
  \quad
  \text{ for } t \geq 0
  \text{ and } N \geq 1
\end{equation}
and
\begin{equation}
  D_{i,j} :=
  W_{i,j}
  \paren{
    \log \frac{c_i c_j}{Q_i Q_j} - \log \frac{c_{i+j}}{Q_{i+j}}
  }
  \geq 0
  \quad
  \text{ for } t > 0
  \text{ and } i,j \geq 1,
\end{equation}
where the $W_{i,j}$ were defined in eq. \eqref{eq:def-Wij} as $W_{i,j}
:= a_{i,j} c_i c_j - b_{i,j} c_{i+j} $, and the time dependence is
implied. Take $T > 0$. Then, for any $N \geq 1$, calculating the time
derivative of $V_N$ from the DCF equations \eqref{eq:DCF} gives
\begin{equation}
  \label{eq:H_thm_finite}
  \begin{split}
    V_N(T) - V_N(0) =
    & - \frac{1}{2} \int_0^T \sum_{i+j \leq N} D_{i,j}(t) \,dt
    \\
    & - \int_0^T \sum_{i=1}^N \sum_{j=N-i+1}^\infty W_{i,j}(t)
    \log\frac{c_i(t)}{Q_i} \,dt ,
  \end{split}
\end{equation}
which can be obtained by a direct calculation after differentiating
$V_N$, as the sum defining it has only a finite number of terms. As
$V$ is finite, $\lim_{N \to \infty} V_N(T) = V(T)$, so the result is
proved if we can show that the right hand side of the above equality
converges to $\int_0^T D_{CF}(t) \,dt$ as $N \to \infty$. To do that,
let us first show that $D_{CF}$ is locally integrable.

Let us find an upper bound for the rightmost term in
\eqref{eq:H_thm_finite}.  It holds that
\begin{align}
  - W_{i,j} \log\frac{c_i}{Q_i}
  & =
  (b_{i,j} c_{i+j} - a_{i,j} c_i c_j)
  \log\frac{c_i}{Q_i}
  \\
  \label{eq:H_thm_1}
  & \leq
  C\, i\, b_{i,j} c_{i+j}
  + a_{i,j} c_i c_j
  \abs{\log\frac{c_i}{Q_i}}
  =: S_{i,j}
\end{align}
for some constant $C \geq 0$ which depends only on $\rho$ and the
$Q_i$, thanks to hypothesis \ref{hyp:Qjk} and the fact that $c_i \leq
\rho$ for all $i$. We have that
\begin{equation}
  \label{eq:H_thm_2}
  \sum_{i,j \geq 1} a_{i,j} c_i(t) c_j(t)
  \abs{\log\frac{c_i(t)}{Q_i}}
  \leq C
  \quad \text{ for all } t \in [0,T]
\end{equation}
for some $C \geq 0$, thanks to lemmas \ref{lem:moment_k+logc},
\ref{lem:moment_k+logQ}, our assumption on moments, and hypotheses
\ref{hyp:growth} and \ref{hyp:Qjk}. Also, using the bound of $b_{i,j}$
in hyp. \ref{hyp:growth},
\begin{equation}
  \sum_{i,j \geq 1} i b_{i,j} c_{i+j}
  =
  \sum_{i=1}^\infty \sum_{j=i+1}^\infty
  c_j i b_{i,j-i}
  =
  \sum_{j=2}^\infty c_j \sum_{i=1}^{j-1}
  i b_{i,j-i}
  \leq
  K \sum_{j=2}^\infty
  c_j j^{\gamma+1},
\end{equation}
which is again uniformly bounded on $[0,T]$. Together with
\eqref{eq:H_thm_2}, this proves that $\sum_{i,j} S_{i,j}$ is uniformly
bounded on $[0,T]$, and in particular that
\begin{equation}
  -\sum_{i=1}^N \sum_{j=N-i+1}^\infty
  W_{i,j}(t)
  \log\frac{c_i(t)}{Q_i}
  \leq C
  \quad
  \text{ for all } t \in [0,T]
\end{equation}
for some $C \geq 0$. Then, from \eqref{eq:H_thm_finite} we deduce
that
\begin{equation}
  \int_0^T \sum_{i+j \leq N} D_{i,j}(t) \,dt \leq C
  \quad \text{ for all } N \geq 1
\end{equation}
for some other constant $C$, as $\abs{V_N(t)}$ is uniformly bounded
for all times $t \geq 0$. Hence, as $\frac{1}{2} \sum_{i+j \leq N}
D_{i,j}(t)$ converges increasingly to $D_{CF}(t)$ as $N \to \infty$,
the monotone convergence theorem shows that $D_{CF}$ is integrable on
$[0,T]$ and that
\begin{equation}
  \frac{1}{2} \int_0^T \sum_{i+j \leq N} D_{i,j}(t) \,dt
  \to
  \int_0^T D_{CF}(t) \,dt
  \quad \text{ when } N \to \infty.
\end{equation}

The previous calculations show that
\begin{equation}
  - W_{i,j}(s) \log\frac{c_i(s)}{Q_i}
  \leq S_{i,j}(s)
\end{equation}
for some $S_{i,j} \geq 0$ such that $\int_0^T \sum_{i,j \geq 1}
S_{i,j}(s) \,ds < +\infty$. Similarly,
\begin{align}
  W_{i,j} \log\frac{c_i}{Q_i}
  & =
  D_{i,j}
  - W_{i,j} \log\frac{c_j}{Q_j}
  + W_{i,j} \log\frac{c_{i+j}}{Q_{i+j}}
  \\
  & \leq
  D_{i,j}
  + S_{j,i}
  + a_{i,j} c_i c_j \log\frac{c_{i+j}}{Q_{i+j}}
  - b_{i,j} c_{i+j} \log\frac{c_{i+j}}{Q_{i+j}}
  \\
  & \leq
  D_{i,j}
  + S_{j,i}
  + C a_{i,j} c_i c_j (i+j)
  + b_{i,j} c_{i+j} \abs{\log\frac{c_{i+j}}{Q_{i+j}}}
  := s_{i,j},
\end{align}
for the same constant $C$ as in \eqref{eq:H_thm_1}. Using our previous
knowledge that both $\sum_{i,j} D_{i,j}$ and $\sum_{i,j} S_{i,j}$ are
integrable on $[0,T]$ and a calculation very similar to the one
carried out before, one can show that $\sum_{i,j} s_{i,j}$ is also
integrable on $[0,T]$. Then, the dominated convergence theorem proves
that the last term in \eqref{eq:H_thm_finite} converges to 0 as $N \to
\infty$, which finishes the proof.

% -------------------------------------------------------------------------
\bibliographystyle{abbrv}
\bibliography{cf,other}

\end{document}